\documentclass[12pt]{article}
\usepackage{epsfig}
\usepackage{graphics}
\usepackage{times}
\begin{document}

\title{Characteristics of solar wind rotation}
\author {K. J. LI$^{1, 3, 4}$,  W.  FENG$^{2}$     \\
\footnotesize{$^{1}$Yunnan Observatories, Chinese Academy of Sciences, Kunming 650011, China}\\
\footnotesize{$^{2}$Research Center of Analysis and Measurement, Kunming University of Science and Technology, Kunming 650093, China}\\
\footnotesize{$^{3}$Center for Astronomical Mega-Science, Chinese Academy of Sciences, Beijing 100012, China} \\
\footnotesize{$^{4}$Key Laboratory of Solar Activity, National Astronomical Observatories, CAS, Beijing 100012, China} \\
}

\date{}
\baselineskip24pt
\maketitle

\begin{abstract}
Over 54 years of hourly mean value of solar wind velocity from  27 Nov. 1963 to  31 Dec. 2017 are used to investigate characteristics of the rotation period of solar wind through auto-correlation analysis. Solar wind of high velocity is found to rotate faster than low-velocity wind, while its rotation rate increases with velocity increasing, but in contrast for solar wind of low velocity, its rotation rate decreases with velocity increasing. Our analysis shows that solar wind of a higher velocity  statistically possesses a faster rotation rate for the entire solar wind. The yearly rotation rate of solar wind velocity does not follow   the Schwable cycle, but it is significantly negatively correlated to yearly sunspot number when it leads  by 3 years. Physical explanations are proposed to these findings.\\
{\bf keywords}  Sun: corona -- Sun: activity -- Sun: atmosphere
\end{abstract}

\section{Introduction}
The solar wind is a supersonic flow of charged  particles, confined by the solar open magnetic field to emanate outward from the solar corona.
Studies on solar wind can provide scientific basis for understanding physical process of space weather, and thus investigations of solar wind are needed. At present observational studies tend to classify solar wind by means of its velocity ($v$) into three categories. (1)Background low-velocity wind,  $ v < 450$ km s$^{-1}$, which mainly emanates from active regions and coronal loops (Harra et al. 2008; Wang et al. 2009; Brooks $\&$ Warren 2011; Morgan 2013; Zangrilli $\&$ Poletto 2016), streamers and pseudo-streamers (Riley $\&$ Luhmann 2012; Crooker et al. 2014; Owens et al. 2014),   and coronal hole boundaries (Nolte et al. 1976; Wang $\&$ Sheeley 1990; Strachan et al. 2002;
Moore et al. 2011; Madjarska et al. 2012; Raouafi et al. 2016). (2)Background high-velocity wind, $450\leq v < 725$ km s$^{-1}$, which streams  principally from large coronal holes. And (3) extreme-high-velocity transient streams, $v\geq 725$ km s$^{-1}$, which generally originate from solar violent eruptions (Marsch 2006a, 2006b; McComas et al. 2007; Zurbuchen 2007; Wang 2012; Deng et al. 2013;  Li, Zhanng¡¡ $\&$ Feng 2016; Li, Zhang $\&$ Feng 2017).
Sixty years of research on solar wind have made great progress, and details have been once given by  Cranmer (2009, 2012), Bruno  $\&$  Carbone (2013),  Owens  $\&$ Forsyth (2013), Cranmer, Gibson $\&$ Riley(2017), and Richardson (2018).

Two main sources of solar wind are though to be  solar active regions and coronal holes, which  usually last for more than one solar rotation (Yin et al. 2007; Shi $\&$ Xie 2014; Xiang, Qu $\&$ Zhai 2014; Yin, et al. 2014; Xiang $\&$ Qu 2016; Gao $\&$ Xu 2016; Xu $\&$ Gao 2016; Badalyan $\&$  Obridko 2018a, 2018b), and rotation should thus be a remarkable attribute.
The solar rotation cycle of about 27.5 days, its 1/3 harmonic period of about 9.1 days  and  1/2 harmonic period of about 13.7 days  were indeed determined in the solar wind speed (Svalgaard $\&$ Wilcox 1975;  Clua de Gonzalez et al. 1993; Nayar et al. 2001;  Katsavrias, Preka-Papadema $\&$  Moussas 2012).
The 1/3 harmonic period is possibly due to the warping of the heliospheric current sheet (Sabbah $\&$ Kudela, 2011; Temmer, Vr$\breve{s}$nak, $\&$ Veronig 2007; Lei et al. 2008; Xie, Shi, $\&$ Xu 2012; Li, Zhang,  $\&$ Feng 2017), and why the  1/2  harmonic period appears is because of the occurrence of two streams per one solar rotation (Fenimore et al. 1978; Gonzalez $\&$ Gonzalez 1987; Mursula $\&$ Zieger 1996; Katsavrias, Preka-Papadema $\&$  Moussas 2012). The rotation period of solar wind display one unique feature, that is,  wave  packet. A series of peaks of statistical significance stand very close to the peak of the rotation period, forming ``wave  packet".
This is due to that several  impulse  streams of solar wind continuously blow, existing for a varying number of solar rotations (Fenimore et al. 1978; Li, Zhang and Feng 2017).

The OMNI 2  records the longest hourly observations of solar wind velocity with time resolution of 0.042 days, and thus this data set is suitable to investigate characteristics of rotation of solar wind. In this case, auto-correlation analysis is carried out in this study, and some interesting results are obtained.

\section{Rotation of solar wind}
\subsection{Data and method}
The data set of the OMNI2 is created as a successor of the original OMNI data set in 2003 at US national space science data center. The set gives the  solar wind plasma and interplanetary magnetic  field parameters, which are observed  near  the  Earth's  orbit by various spacecraft (King $\&$  Papitashvili 2005; Kasper et al. 2006).
Hourly mean value  of solar wind velocity ($v$ in $km s^{-1}$) is used in this study, which is downloaded from the  OMNIWeb\footnote{http://omniweb.sci.gsfc.nasa.gov/html/ow$_{-}$data.html}. In this study the unit of solar wind velocity is always to be $km s^{-1}$, and thus it should be $km s^{-1}$ if unit of velocity is not given next.
Figure 1 shows hourly mean value of solar wind velocity from 12 o'clock of 27 Nov. 1963 to 23 o'clock of 31 Dec. 2017, spanning over 54 years.
During this time interval hourly solar wind velocity has 354654 measurement records, making up  $74.8\%$ of the total 474204, and it is a serious uneven big data set.
Figure 2 shows yearly number of observation records of hourly average velocity. The minimum and maximum values of this time series are 156 and 1189, respectively. That is, the velocity interval of the entire original time series is from 156 (the lower limit) and 1189 (the upper limit).

\begin{figure}
\hskip -1.0 cm
\includegraphics[width=12. cm, height=12. cm]{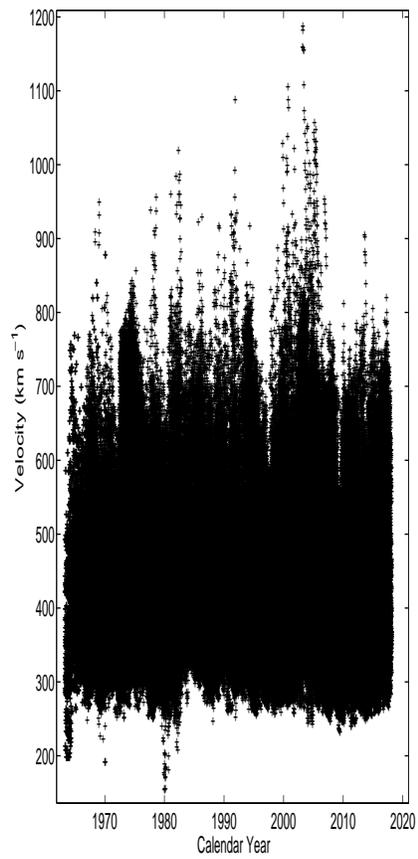}
\caption{
Hourly mean value (pluses) of solar wind velocity from 12 o'clock of 27 Nov. 1963 to 23 o'clock of 31 Dec. 2017.
}
\label{fig: figure1}
\end{figure}

\begin{figure}
\hskip -1.0 cm
\includegraphics[width=12. cm, height=12. cm]{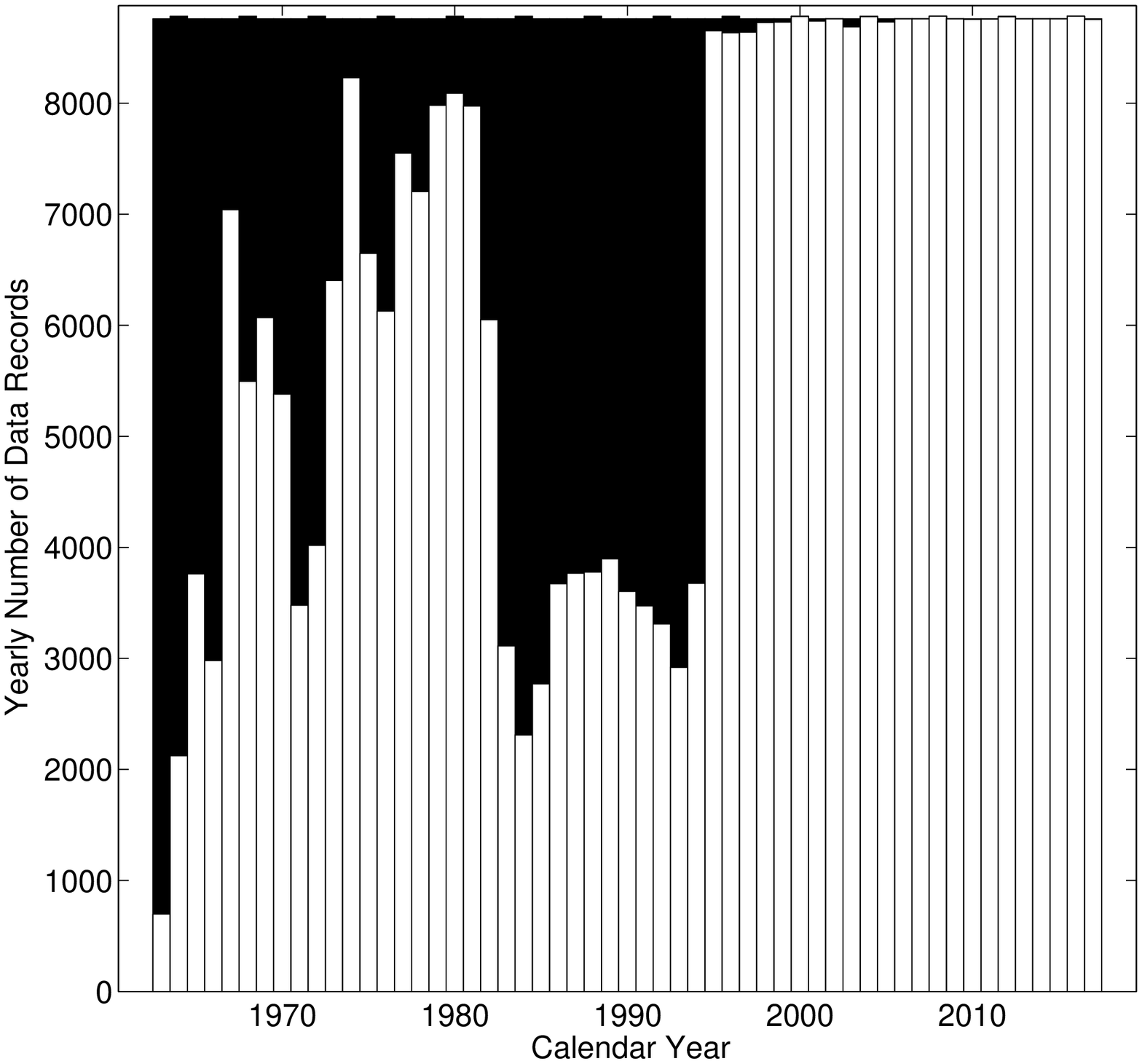}
\caption{Annual number of observation records for hourly solar wind velocity (the white histogram). The black histogram  above the white histogram shows yearly number of no records.
}
\label{fig: figure2}
\end{figure}

After the year of 1994,  wind velocity is measured in almost all hours, but no records frequently occur in and before the year.
For such an uneven big data (474204 records),  general methods (program softwares) of the Fourier transform cannot smoothly run in a microcomputer, and  an out-of-memory error should occur.  Here the auto-correlation analysis of a time series is carried out to show  auto-correlation coefficient of the series varying with relative phase shifts. Here, the so-called   ''relative phase shifts"  actually refer to phase shifts of a time series moving related to the other of two exactly same series.
If no measurement of wind velocity in a certain hour is obtained, then the hour is ignored and does not participate in calculation of auto-correlation coefficient.
Li et al (2012) and Xu and Gao (2016) gave a detail description to the analysis method. This study aims to investigate characteristics of rotation of solar wind, and the method is simple but practical. In this study, what we are measuring is recurrence period.

\subsection{Rotation Period of solar wind}
As mentioned above, solar wind  velocity can be quantitatively divided  into  the classifications of low- and high-velocity wind.
Figure 3 shows auto-correlation coefficient of three kinds of solar wind: all-, low- and high-velocity wind, varying with relative phase shifts, and  their rotation periods are 27.13, 27.21, and 26.83 days, respectively. High-velocity wind seems to rotate faster than low-velocity wind. The 13.96-day period  appears  for the all-velocity wind. However for the high-velocity wind, multiple  periods, 11.67, 13.62 and  16.00 (days) seem to form
the so-called ''wave packet". Next, rotation periods of time series of wind velocity will be determined always by means of auto-correlation analysis, and thus the analysis method is no longer mentioned. In this study, what we are measuring is recurrence period.

\begin{figure}
%\hskip -5.0 cm
\hskip 2.0 cm
\includegraphics[width=12. cm, height=12. cm]{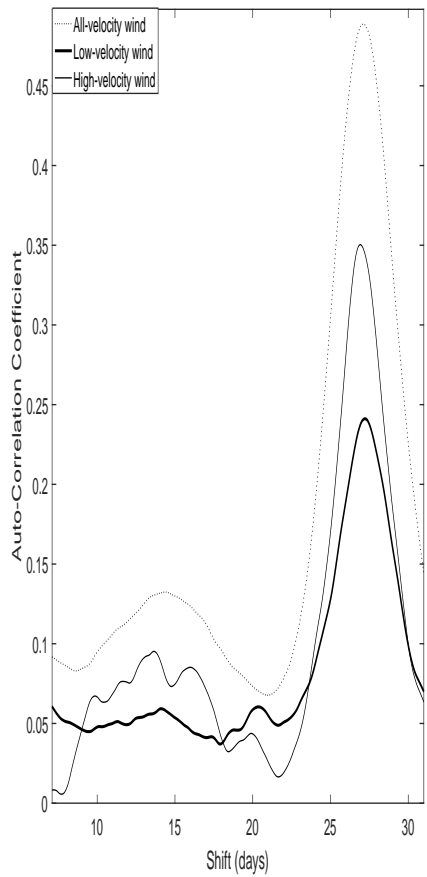}
\caption{Auto-correlation coefficient of all- (the dotted line), low- (the thick solid line), and high-velocity wind (the thin solid line), varying with relative phase shifts.
}
\label{fig: figure3}
\end{figure}

As the Figure 2 of Li, Zhanng $\&$ Feng (2106) shows, the probability  density distribution of solar wind velocity assumes the $\Gamma$-function shape, and the most probable velocity is about 373. There are very few measurements at the two ends of the distribution for velocities  which are larger 700 or smaller than 275, and
less data will affect the determination of rotation period.
Cumulative intervals of velocities may alleviate this problem. Here, ``cumulative intervals" means that one new time series of wind velocity is gotten by adding data of velocity to the previous time series through extending velocity interval of the previous time series (one end of velocity interval of the previous time series is extended, while the other end is fixed).
Next we will investigate rotation periods of different velocity intervals of solar wind in details in two ways, and the first one is cumulative intervals of velocities.

Setting the upper limit of velocity interval of solar wind  in turn to be 300, 400, 500, and so on (velocity interval increases by 100 each time),
with the lower limit always to be the minimum velocity of the whole original series of solar wind, we can obtain a few new time series of velocity, and they  are the time series of $v\leq 300$, that of $v\leq 400$, that of $v\leq 500$, and so on, correspondingly.
Figure 4 shows rotation periods of these time series, where their upper limits  are marked as abscissa values of  corresponding periods.
Also shown in the figure are rotation periods of the time series of $v\geq 600$, that of $v\geq 500$, that of $v\geq 400$, and so on (velocity interval increases by 100 each time), where their lower limits are marked as abscissa values, and upper limit of these series is always to be the maximum of the entire original series.

\begin{figure}
\hskip 2.0 cm
\includegraphics[width=12. cm, height=12. cm]{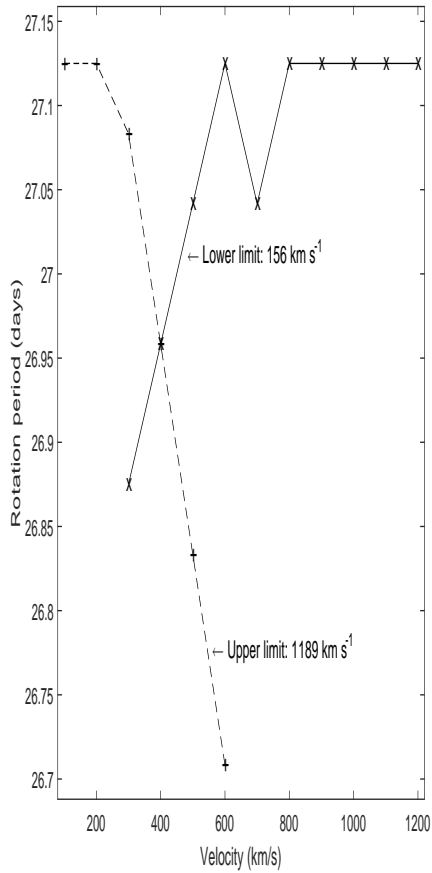}
\caption{Crosses in the solid line: rotation period of a wind-velocity series varying with the maximum velocity of the used series, and pluses in the dashed line: rotation period of a wind-velocity series varying with the minimum velocity of the used series.
Each cross corresponds to a wind-velocity series, and the series is produced by deleting those velocity measurements which are larger than the set maximum velocity from the original all-velocity series. The maximum velocity of the first series is set to be 300, that of the second is 400, that of the third is 500, and so on.
Each plus corresponds to a wind-velocity series, and the series is produced by deleting those velocity measurements which are smaller than the set minimum velocity from the original all-velocity series. The minimum velocity  of the first series is set to be 600, that of the second is 500, that of the third is 400, and so on.
}
\label{fig: figure4}
\end{figure}

As Figure 4 shows, rotation period seems  to increase with wind velocity increasing for the low-velocity wind (please see the leftmost 2 $\sim$ 3 crosses), but for the high-velocity wind it decreases with wind velocity increasing (please see the rightmost 2 $\sim$ 3 pluses). The figure clearly displays that high-velocity wind  rotates faster than low-velocity wind.

Next we will refine the work presented in Figure 4.
Figure 5 shows rotation periods of the time series of $v\geq 300$, that of $v\geq 305$, that of $v\geq 310$, and so on (velocity interval increases by 5 each time), where their upper limits  are marked as abscissa values of  corresponding periods, and lower limit of these series is always to be the minimum of the entire original series.
A linear regression is carried out for those points whose abscissas are smaller than 450  in the figure (namely for low-velocity wind), and
the correlation coefficient ($CC$) is 0.6752 (positive correlation), which is statistically significant at the $99.9\%$ confidence level.
Therefore for the low-velocity wind, rotation period increase overall with wind velocity increasing.
A linear regression is also carried out for those points whose abscissas are smaller 720 but not smaller than 450  in the figure (namely for high-velocity wind), and
the correlation coefficient ($CC$) is -0.6412 (negative correlation), which is statistically significant at the $99.9\%$ confidence level.
Therefore rotation period decreases generally with wind velocity increasing for the high-velocity wind.

\begin{figure}
\hskip 2.0 cm
\includegraphics[width=12. cm, height=12. cm]{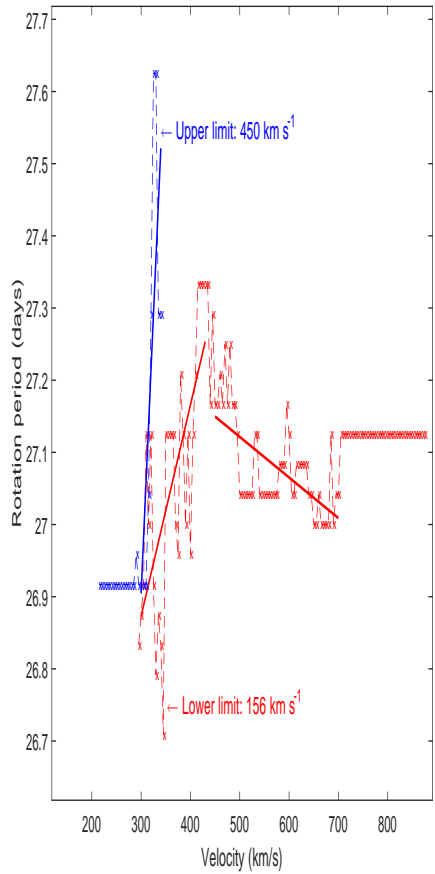}
\caption{Blue crosses in the dashed line: rotation period of a wind-velocity series varying with the minimum velocity of the used series, and the blue solid line is the regression line of the rightmost 9 blue points. Red crosses in the dashed line: rotation period of a wind-velocity series varying with the maximum velocity of the used series, and the red solid lines are the corresponding regression lines of 27 red points (crosses) and  51 red points.
Each blue cross corresponds to a wind-velocity series, and upper limit of all series is always set to be 450, but lower limit is set in turn to be 350,  345, 340, and so on.
Each red cross corresponds to a wind-velocity series, and  lower limit of all series is always set to be the minimum velocity of the original all-velocity wind series, but upper limit is set in turn to be 300, 305, 310, and so on.
}
\label{fig: figure5}
\end{figure}

Also shown in Figure 5 are rotation periods of the time series of
$350\leq v < 450$,  the series of  $345\leq v < 450$, the series of $340\leq v < 450$,  and so on, where their lower limits  are marked as abscissa values of  corresponding periods.
Lower limit of velocity interval is extended by 5 each time, and upper limit of these series is fixed to be
the maximum velocity of low-velocity wind (450). As the figure displays, rotation period does no longer change with the increase of  velocity interval (extending lower limit) after about 310.
For the rightmost 9 points of these obtained rotation periods, a linear regression can give a good description with the correlation coefficient ($CC$) being 0.7801 (positive correlation),
which is significant at the $98\%$ confidence level. Therefore,
the figure also shows that, rotation period increases with wind velocity increasing for the low-velocity wind, but for the high-velocity wind it seems to decrease with wind velocity increasing.

Similarly, rotation periods are determined for these time series of wind velocity, $v \geq 660$, $v \geq 655$, $v \geq 650$, and so on (lower limit is extended by 5 at one time).
The obtained rotation periods for these series of wind velocity are given in Figure 6  as ordinate values, while their abscissas are marked as their corresponding minimum velocities (lower limits).
For the high-velocity wind (the rightmost 43 plus-points), rotation period increases overall with the decrease of wind velocity, and a linear regression can well describe the relation between rotation periods and corresponding minimum velocities with $CC$ is -0.7420 (negative correlation), which is statistically significant at the $99.9\%$ confidence level.

\begin{figure}
\hskip 2.0 cm
%\hskip -5.0 cm
\includegraphics[width=12. cm, height=12. cm]{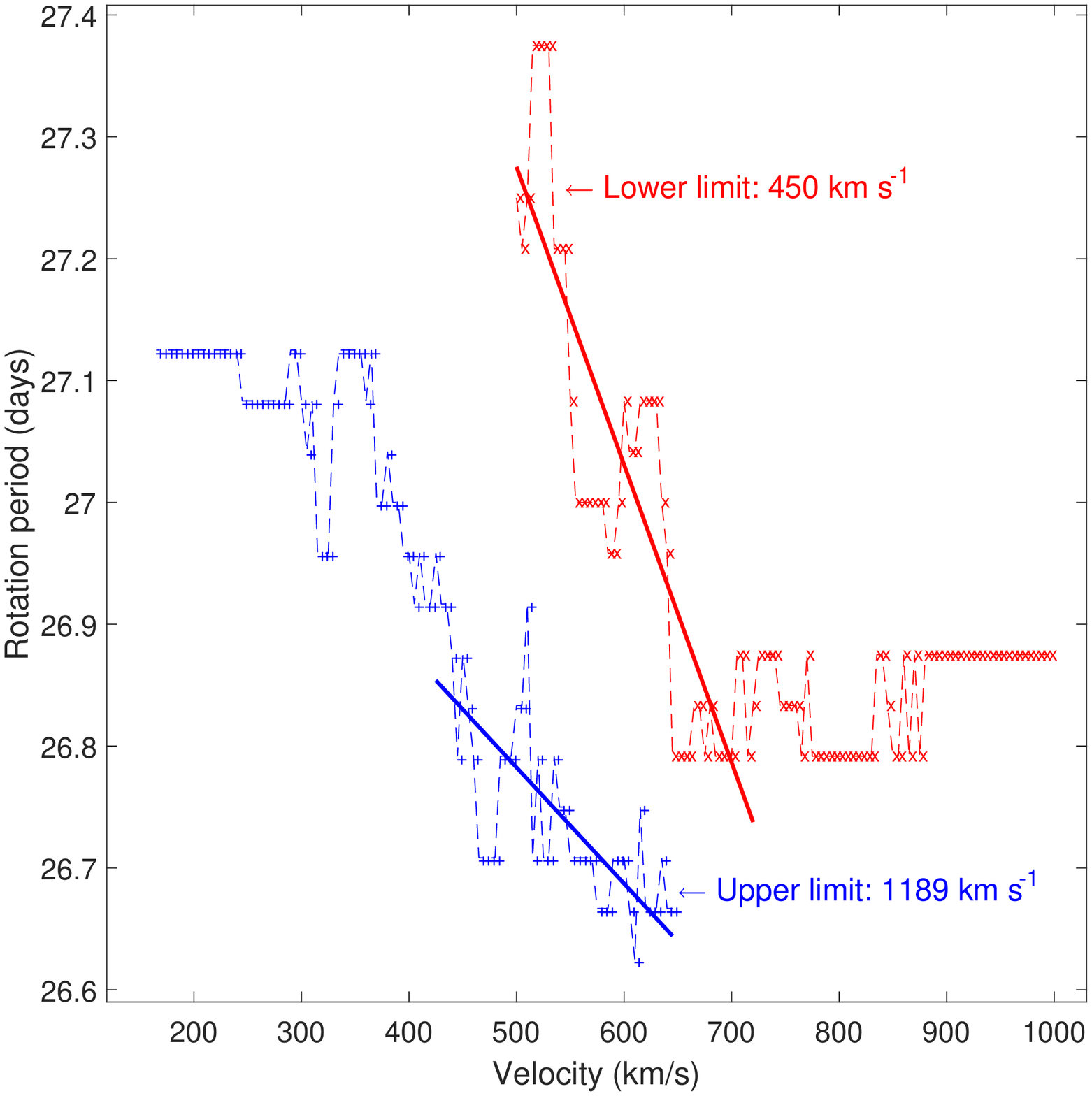}
\caption{Pluses in the blue dashed line: rotation period of a wind-velocity series varying with the minimum velocity of the used series, and the blue  solid line is the corresponding regression line of the rightmost 43 points marked by pluses.
Crosses  in the red dashed line: rotation period of a wind-velocity series varying with the maximum velocity of the used series, and
the  red solid line is the corresponding regression line of  the leftmost 45 points marked by crosses.
Each plus corresponds to a wind-velocity series, and upper limit of all series is always set to be the largest velocity of the original all-velocity series, but  lower limit is set in turn to be 660,  655, 650, and so on.
Each cross corresponds to a wind-velocity series, and lower limit of all series is always set to be 450, but upper limit is set in turn to be 500,  505, 510, and so on.
}
\label{fig: figure6}
\end{figure}

Similarly, rotation periods are determined for the time series of wind velocity, $450\leq v \leq 500$, $450\leq v \leq 505$, $450\leq v \leq 510$, and so on. Upper limit of these series is extended by 5 at one time, and lower limit is fixed to be the minimum velocity of high-velocity wind.
The obtained rotation periods for these series of wind velocity are given in Figure 6  as ordinate values, while their abscissas are marked as their corresponding upper limits.
For the high-velocity wind (the leftmost 45 cross-points), rotation period increase overall with the decrease of wind velocity, and a linear regression can well describe the relation between the two with $CC$ is -0.8734 (negative correlation), which is statistically significant at the $99.9\%$ confidence level.

Finally, rotation periods are determined in the other way for a few time series of wind velocity, whose intervals are fixed always to be 100. Middle value of velocity  interval of these series, $v_{mid}$ is set in turn to be 250,  255, 260, and so on ($v_{mid}$ increases by 5 at one time). Therefore, these wind-velocity series are set in turn to be the series of 250-50 to 250+50, the series of
255-50 to 255+50, the series of 260-50 to 260+50, and so on.
Here, $v_{mid}$ of a series does not mean the average or median value of all velocities  of the series.
Figure 7 displays rotation periods, varying with $v_{mid}$s which are marked as the correspondingly abscissa values.
For the leftmost 38 series whose middle values ($v_{mid}$) of velocity interval are 250 to 435, a linear regression can well describe the relation between
middle value of interval of a series and rotation period of the series,  with $CC$ being 0.686 (positive correlation), which is statistically significant at the $99.9\%$ confidence level.
When $v_{mid}=435$, the corresponding series is for solar velocity ranging from 385 to 485, most of which  belong to low-velocity wind.
For the rightmost 48 series whose middle values ($v_{mid}$) of velocity interval are 465 to 700, a linear regression can well describe the relation between the two,
with $CC$ being  -0.416 (negative correlation), which is statistically significant at the $99.4\%$ confidence level.
For the low-velocity wind, rotation period increase overall with the increase of wind velocity, but for the high-velocity wind, rotation period decrease overall with the increase of wind velocity.

\begin{figure}
\hskip 2.0 cm
%\hskip -5.0 cm
\includegraphics[width=12. cm, height=12. cm]{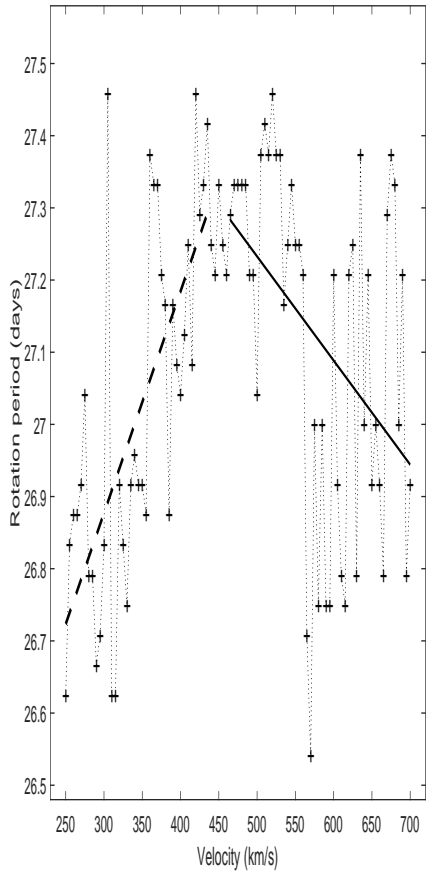}
\caption{Pluses in the dotted line: rotation period of a wind-velocity series varying with the middle velocity ($v_{mid}$) of the fixed interval (100) of the used series, and the dashed and solid lines are the corresponding regression lines of the leftmost 38 points  and the rightmost 48 points.
Each plus corresponds to a wind-velocity series of the fixed interval, and its speed interval is $v_{mid}-50$ to $v_{mid}+50$.  Middle velocity, $v_{mid}$ is set in turn to be 250,  255, 260, and so on. That is, wind-velocity series are set in turn to be the series of 250-50 to 250+50, the series of
255-50 to 255+50, the series of 260-50 to 260+50, and so on.
}
\label{fig: figure7}
\end{figure}

\subsection{Variation of rotation period of solar wind within sunspot cycles}
The whole time series of the original all-velocity wind is cut into series within each calendar year, and then these series are utilized to determine rotation period in each calendar year by means of the auto-correlation analysis method mentioned above. Figure 8 shows the obtained rotation period for wind velocity in each calendar year.
Rotation period of solar wind velocity in one year is highly time-varying.

\begin{figure}
\hskip -1.0 cm
\includegraphics[width=12. cm, height=12. cm]{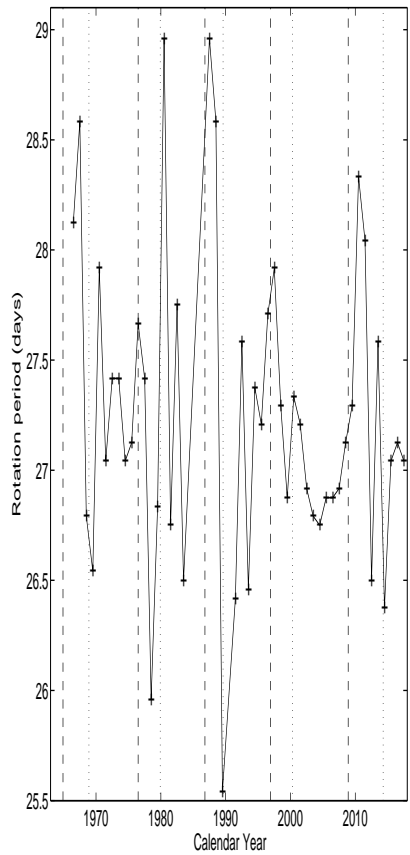}
\caption{Rotation period (pluses) of solar wind velocity within one calendar year. The vertical dashed/dotted lines  display the minimum/maximum times of sunspot cycles.
}
\label{fig: figure8}
\end{figure}

Yearly mean sunspot numbers in years of 1964 to 2017,  which is downloaded from the web site of the Solar Influences Data Analysis Center (http://sidc.oma.be) and shown in Figure 9, is  used to take a cross-correlation analysis with  rotation period of solar wind in one year, which is shown also in Figure 9. Resultantly Figure 10 shows $CC$ varying with their relative phase shifts with phases of the latter leading the former given positive values, that is, positive phase-shift values mean that yearly rotation period  of solar wind velocity is in advance of the yearly sunspot number.
When yearly rotation period  of solar wind velocity leads yearly sunspot number by 3 years, $CC$ is 0.3192, which is significant at the $96\%$ confidence level. It is known that daily velocity of solar wind leads daily sunspot number by about 3 years (see the Figure 3 of Li, Zhanng and Feng (2016)), therefore yearly mean velocity of solar wind should possibly in phase with yearly rotation period  of solar wind velocity.

\begin{figure}
\hskip -1.0 cm
\includegraphics[width=12. cm, height=12. cm]{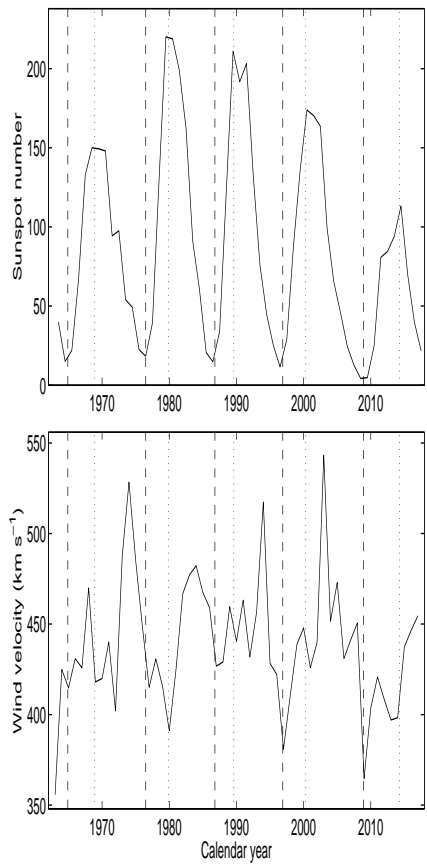}
\caption{Top panel: yearly sunspot number from 1964 to 2017. Bottom panel: yearly mean velocity of solar wind from 1964 to 2017.
}
\label{fig: figure9}
\end{figure}

\begin{figure}
\hskip -1.0 cm
\includegraphics[width=12. cm, height=12. cm]{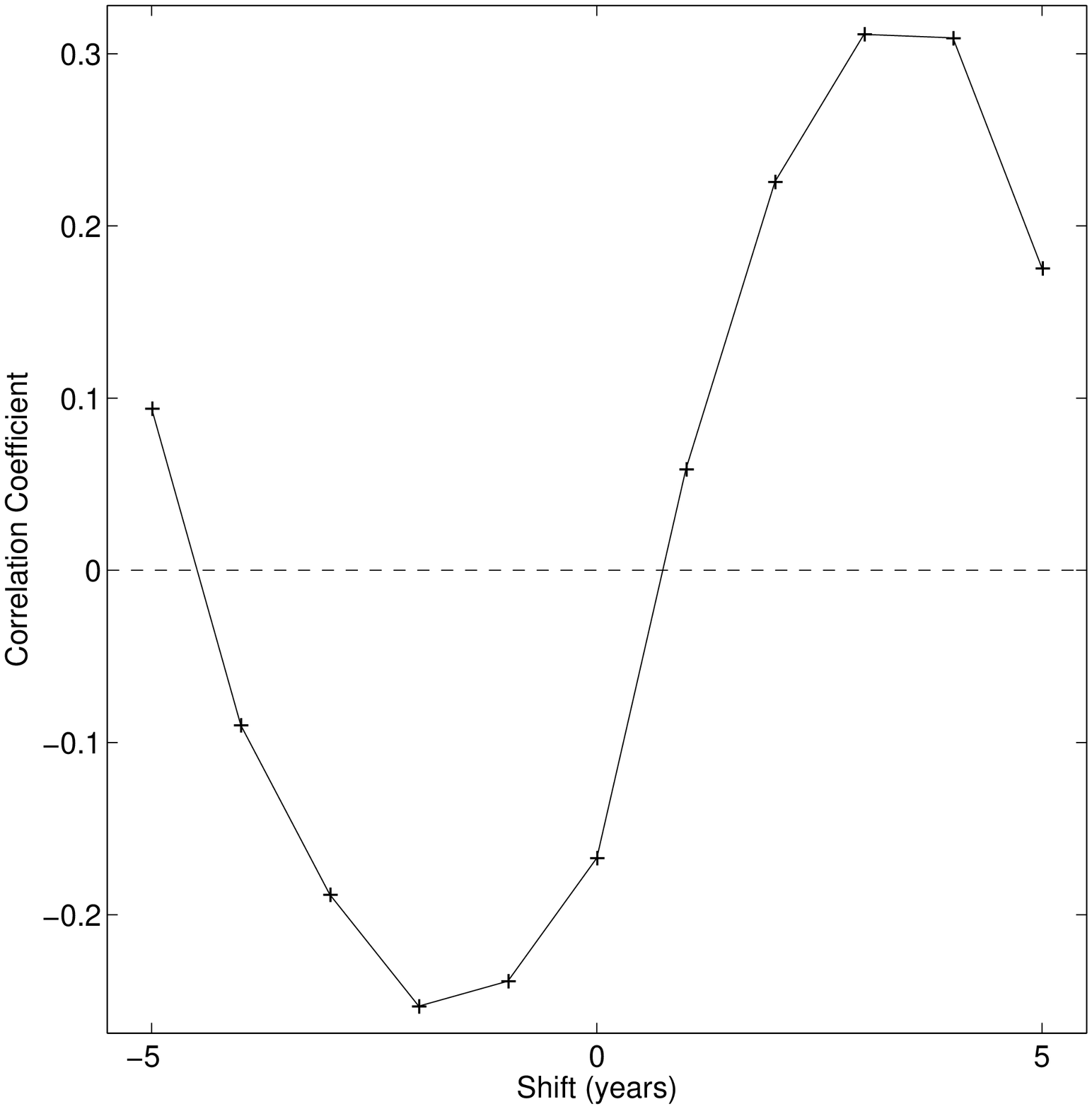}
\caption{Cross-correlation coefficient (pluses) between  rotation period  of solar wind velocity in one year and  yearly sunspot number, varying with their relative phase shifts with backward shifts given minus values.
}
\label{fig: figure10}
\end{figure}

Yearly mean velocity of solar wind in years of 1964 to 2017 is  calculated, which is shown in Figure 9, and it is  utilized to take a cross-correlation analysis with yearly rotation period of solar wind. Resultantly Figure 11 shows $CC$ varying with their relative phase shifts, and
positive phase-shift values  mean that yearly mean velocity of solar wind is delayed in respect to yearly rotation period of solar wind.
$CC$ reaches  its minimum, -0.2961 at no relative shift, which is significant at the $96\%$ confidence level.  It  means that, yearly mean velocity of solar wind temporally fluctuates in the opposite way as yearly rotation period of solar wind does, and
thus higher velocity of solar wind should statistically possess shorter rotation period.

\begin{figure}
\hskip -1.0 cm
\includegraphics[width=12. cm, height=12. cm]{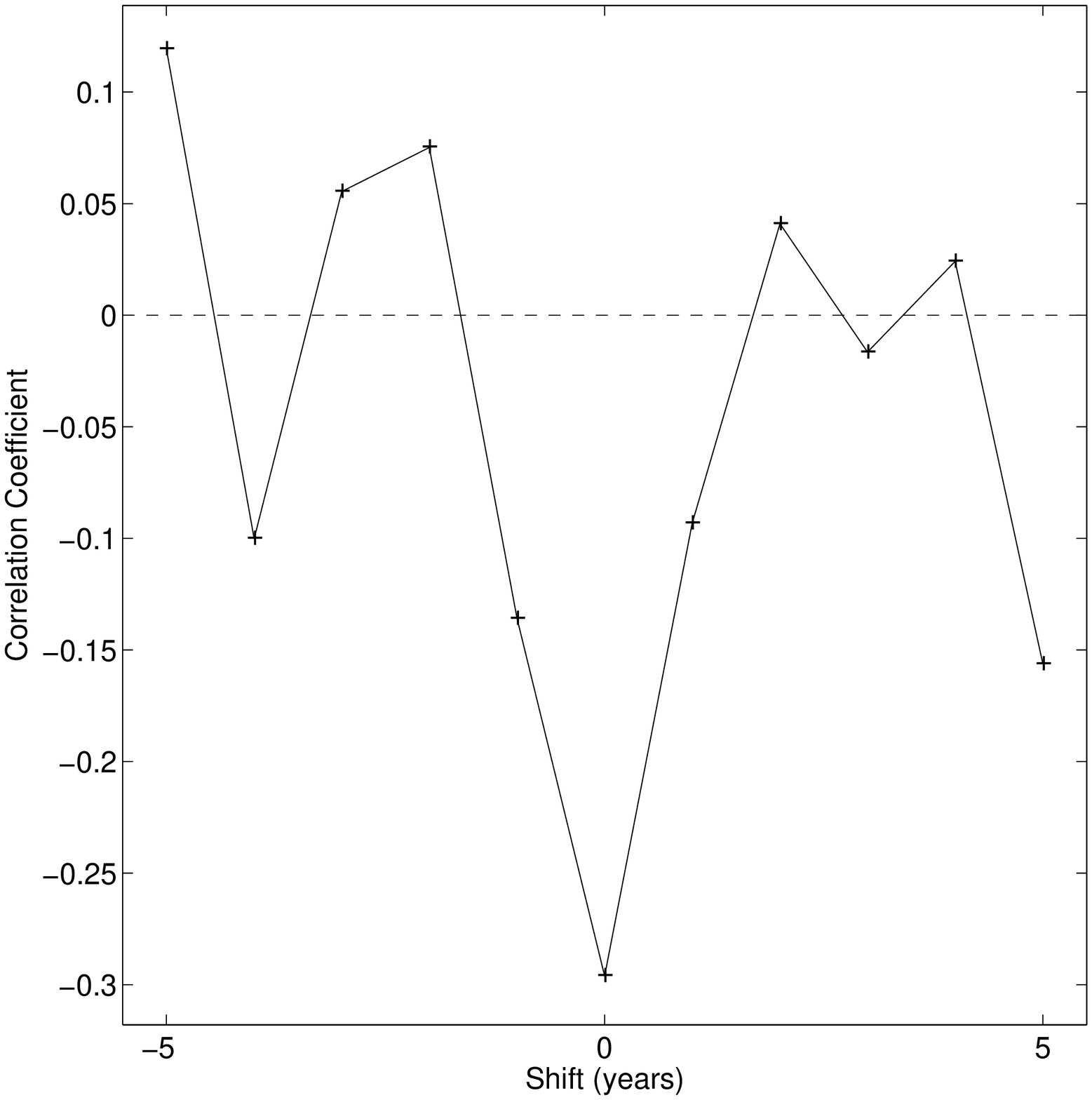}
\caption{Cross-correlation coefficient (pluses) of  yearly rotation period  of solar wind velocity with yearly mean velocity of solar wind, varying with their relative phase shifts with backward shifts given minus values.
}
\label{fig: figure11}
\end{figure}

The National Geophysical Data Center (NGDC) of USA gives the minimum ($t_{min}$) and maximum ($t_{max}$) times of a sunspot cycle
%$\sf{https://www.ngdc.noaa.gov/stp/space-weather/solar-data/solar-indices/sunspot-numbers/cycle-data/}$,
$https://www.ngdc.noaa.gov/stp/space-weather/solar-data/solar-indices/sunspot-numbers/cycle-data/$,
and accordingly we can construct the time series of wind velocity
in the minimum time of a sunspot cycle from  $t_{min}-1.5$  to $t_{min}+1.0$ in years and that in the maximum time of a sunspot cycle from  $t_{max}-1.0$  to $t_{max}+1.5$ in years,
respectively for all-, low-, and high-velocity wind. Then we can determine rotation periods of these series by means of auto-correlation analysis method,
which are shown correspondingly in Figures 12 to 14.

Rotation period in the minimum time of a sunspot cycle is larger than that in the maximum time  in 4 cycles of the total 5 for these three kinds of solar wind.
Linear regression indicates that the two are positively correlated for all three kinds, but correlation coefficients are all of statistical insignificance.

\begin{figure}
\hskip -1.0 cm
\includegraphics[width=12. cm, height=12. cm]{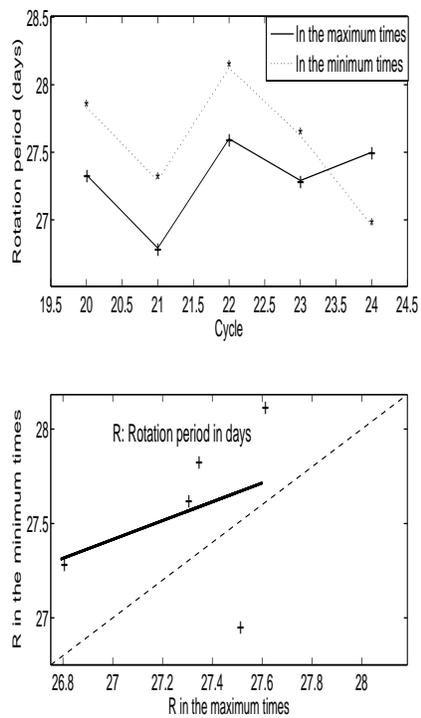}
\caption{Top panel: rotation period in the minimum (asterisks in a dashed line) and maximum (pluses in a solid line) time of sunspot cycles 20 to 24  for all-velocity solar wind.
Bottom panel: relation (pluses) of rotation period ($R$) in the minimum  times $vs$ that in the maximums for all-velocity solar wind. The solid line is its linear fitting, and the dashed line shows identical abscissa and ordinate values.
}
\label{fig: figure12}
\end{figure}

\begin{figure}
\hskip -1.0 cm
\includegraphics[width=12. cm, height=12. cm]{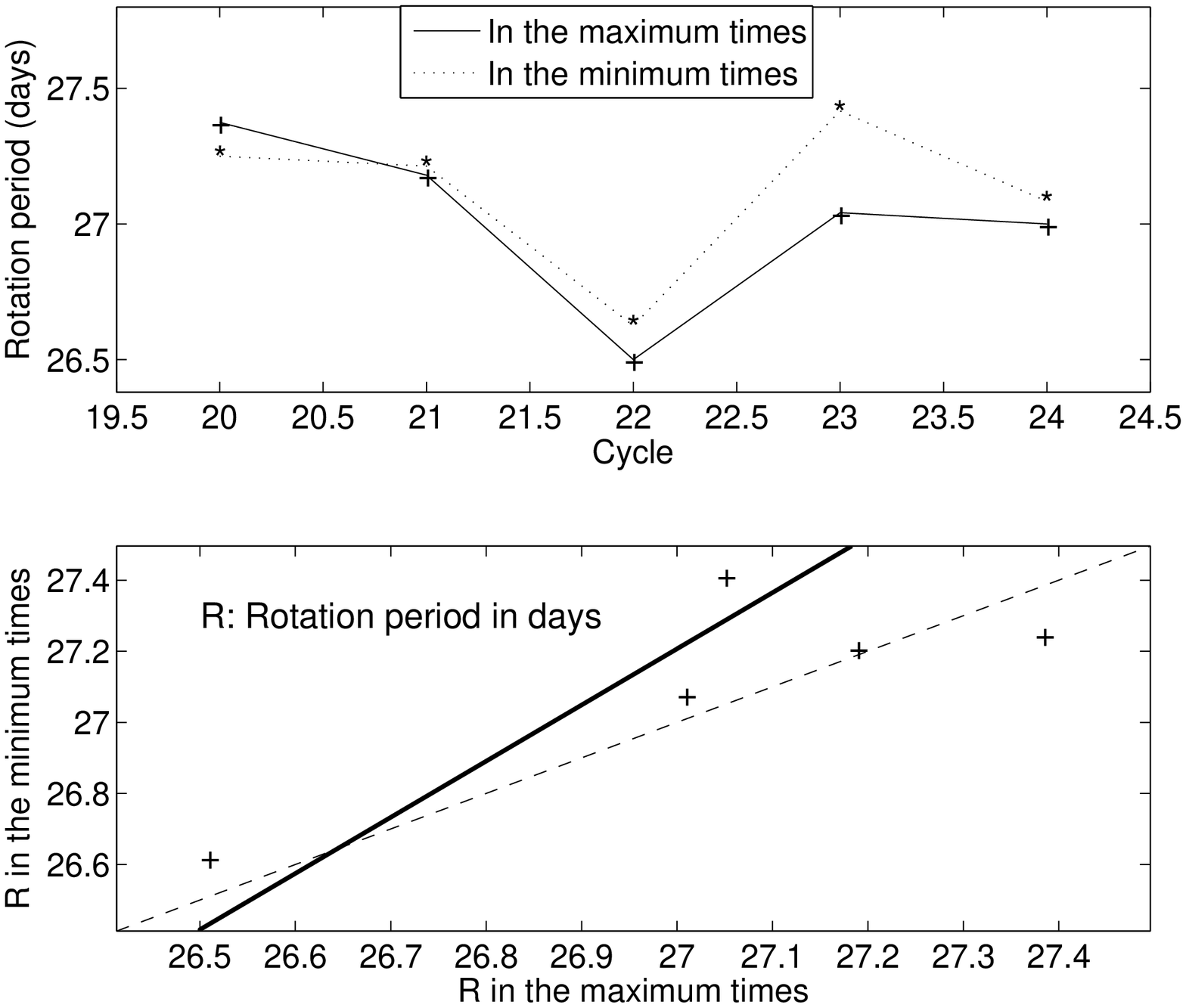}
\caption{The same as Figure 12, but for low-velocity solar wind instead of all-velocity solar wind.
}
\label{fig: figure13}
\end{figure}

\begin{figure}
\hskip -1.0 cm
\includegraphics[width=12. cm, height=12. cm]{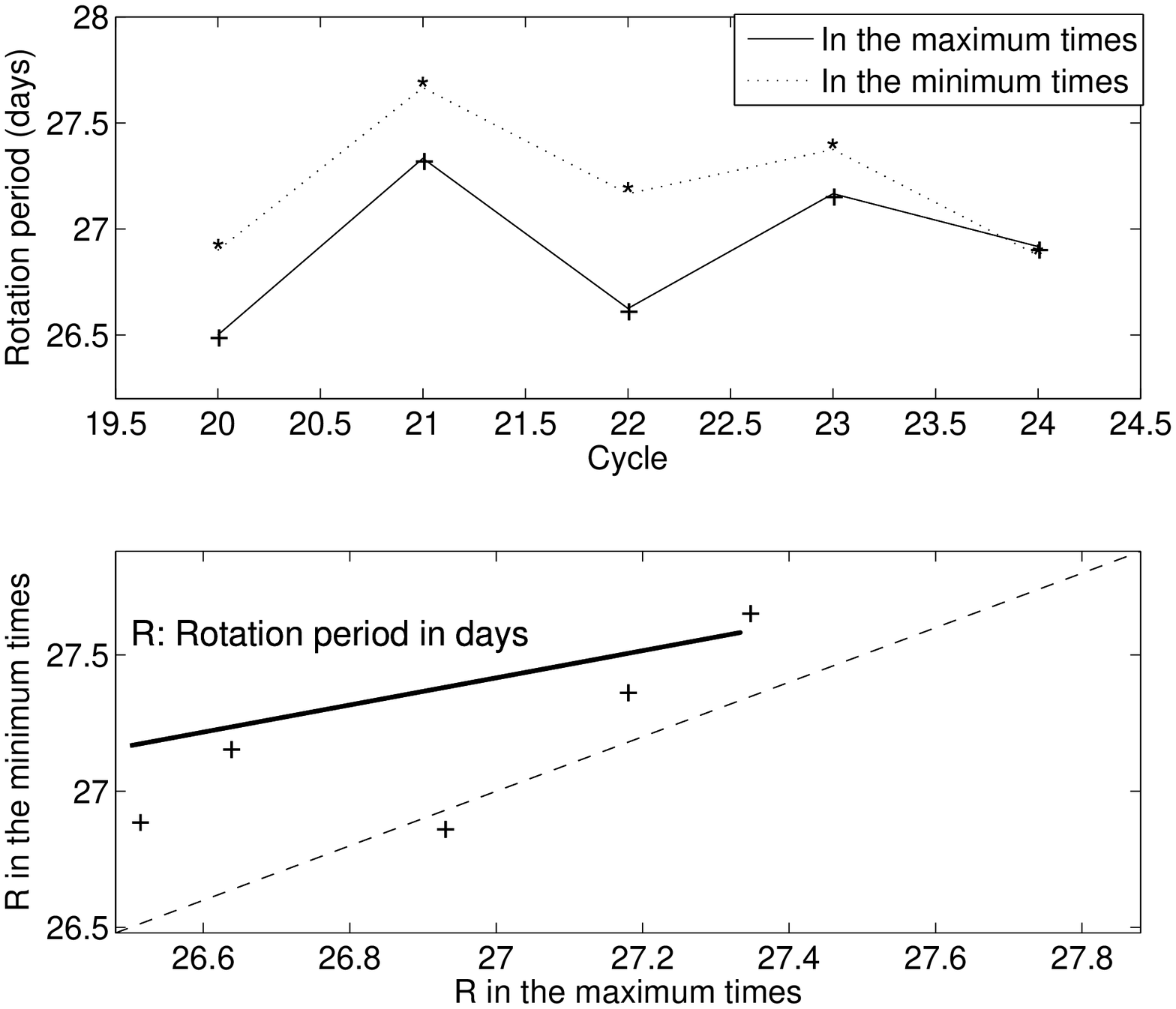}
\caption{The same as Figure 12, but for high-velocity solar wind instead of all-velocity solar wind.
}
\label{fig: figure14}
\end{figure}

\section{Conclusions and Discussion}
The OMNI 2 data set records hourly mean value of solar wind velocity, which is suitable to exactly investigate the variation of the solar rotation period,
and  used here are the data from  12 o'clock of 27 Nov. 1963 to 23 o'clock of 31 Dec. 2017.
Auto-correlation analyses are carried out respectively to the all-, low-, and high-velocity wind at the time interval, and  results are obtained as the followings.

(1) The rotation period for all-, low- and high-velocity wind is 27.13, 27.21, and 26.83 days, correspondingly. Besides the rotation period, the one-second rotation period,
13.96 days is  found to exist  for the all-velocity wind, but for the high-velocity wind, multiple peaks appear around the the one-second rotation period to seemingly form a
 ``wave packet" (Fenimore et al. 1978). However peak CCs of the wave packet are very small, and peaks are fairly gentle, therefore it is not sure that such a wave packet is a true phenomenon, or  it is caused by the serious data missing for the fast wind.
High-velocity wind is determined to rotate faster than low-velocity wind, and this result was once obtained by Li, Zhang, and Feng (2017). They used the Lomb \,¨C\, Scargle periodogram method to analyze daily mean velocity of solar wind with time resolution being  1 day.

Sunspot regions rotate slower than mid-latitude CHs at latitudes above $15^{0}$ (Insley, Moore, Harrison 1995). Recently, sunspots are found to rotate slower than the coronal holes at latitudes higher than $35^{0} - 40^{0}$ (Bagashvili et al. 2017).
Some researches also show that coronal structures rotate faster than sunspot regions (Waldmeier 1950; Cooper $\&$ Billing 18962; Vats et al. 2001; Badalyan 2009).
That is, the main  (or a considerable part of) sources of low-velocity wind  rotate slower than the main sources of high-velocity wind.
Although rotation period of solar wind is found here to vary with wind velocity, but such the variation is slight, and rotation period of any wind velocity is close to (or not far away from) the rotation period of solar magnetic structures (around 27 days), although solar wind is prominently radially accelerated.

CHs and active regions of sunspots may be correspondingly regarded as ``emitters" of some high-  and  low-velocity  wind,
and some solar wind is launched from these rotating emitters.
It is suggested here that it is emitters' rotating that promotes the solar wind to rotate, and thus
rotation rates of these emitters somewhat reflects frequency of that solar wind is recorded and rotation of some solar wind.
This is inferred to be the possible reason why low-velocity wind  rotate slower than  high-velocity wind, and
perhaps future work will be necessary to have a more precise picture of this phenomenon.
Solar wind of low- and high-velocity rotates  with different periods, and this is favourable to the formation of co-rotating interaction regions (CIR), but of course, it is not guaranteed that different rotation rates can certainly cause the formation of CIR.

(2) Rotation period increases with wind velocity increasing for the low-velocity wind, but for the high-velocity wind it decreases with wind velocity increasing.
However, for solar wind as a whole, a higher velocity of solar wind should statistically possess a shorter rotation period.
Long-life large sunspots rotate faster than ordinary sunspots (Newton and Nunn 1951; Stix 1989). A considerable part of low-velocity wind blows out of sunspot regions, and solar wind from long-life large sunspots is suggested to relatively have larger velocities than that from ordinary sunspot regions, and this is inferred to be the reason why rotation period of low-velocity wind increases with wind velocity increasing.
The rotation of CHs is found to increase from the middle latitudes toward the poles  (Navarro-Peralta $\&$ Sanchez-Ibarra  1994),
and meanwhile high-velocity wind is confirmed to dominate at higher latitudes by Ulysses (Pillips et al 1994), and this is inferred to be the reason  why rotation period decreases with wind velocity increasing  for the high-velocity wind.

(3) Rotation period in the minimum time of a sunspot cycle is larger than that in the maximum time  in 4 cycles of the total 5 for all three kinds of solar wind,  implying that solar wind should tend to rotate faster around the maximum time  than that around  the minimum time of a sunspot cycle, and the two are positively correlated, implying that if CHs rotate fast around the minimum time of a cycle, then sunspots should tend to rotate fast at the maximum time of the cycle.
However this trend is statistically insignificant. By the way, data missing should affect the accurate determination of rotation period, especially for the fast wind.
Rotation period of solar wind velocity of one year is highly time-varying, and it is hardly found to be related with sunspot cycles. No obvious cyclic variation  has been found in  the rotation of the green solar corona by Rybak (1994) and in the rotation of the F10.7 corona  by Li et al. (2012).
In cycle 21, the yearly number of polar and mid-latitude coronal holes (CHs) peaks  about-3-year before the ending time of the cycles (Insley, Moore, Harrison 1995).
The more CHs are, the more high-velocity wind events which are produced by the CHs are, and thus the larger the mean velocity of solar wind is. This is why solar wind velocity within a sunspot cycle usually peaks about-3-year before the minimum time of the cycle (see Figures 9 and 10).
So rotation rate of solar wind should peak about-3-year before the minimum time, and this is why rotation period of solar wind is positively correlated with SSN when wind velocity leads SSN by about 3 years. The Figure 3 of Li, Zhanng, Feng (2016) also shows that solar wind velocity should lead SSN by about 3 years.

For both low- and high-velocity wind, the lack of data prevents from determining the rotation cycle of a short data span, and this is the reason why we determine rotation periods of cumulative data spans instead in this study.

Here the above results are attempted to statistically explain through  original sources of solar wind, but the actually measured wind velocity may go through  many physical  processes when solar wind propagates from the origin to the measurement point, including accelerating, interacting with the interplanetary medium or interacting between wind of different velocities, and so on (Elliott et al 2012; Sanchez-Diaz, et al 2016; Cranmer, Gibson $\&$ Riley 2017).
It is likely that the  obtained statistical results  will be violated  by these known and/or unknown processes. However, these processes are believed to
approximately equivalently affect rotation of high- and low- velocity wind, because rotation periods measured at Lagragian point L1, namely,
recurrence periods of given solar wind structures
 are close to these measured in the low corona, although
solar wind undergos obvious acceleration from subsonic to supersonic speed. Then our explanation seems  plausible, and
the original sources of wind are likely to influence the rotation of different solar wind streams.

{\bf acknowledgments:}
We thank the anonymous referee for careful reading of the manuscript and constructive comments which improved the original version of the manuscript.
We acknowledge use of NASA/GSFC's Space Physics Data Facility's OMNIWeb  service, and OMNI data.
This work is supported by  the
National Natural Science Foundation of China (11973085, 11903077, 11633008, and 11573065), the Yunling-Scholar Project, and the national project for large scale scientific facilities.

\clearpage

\end{document}